# Detection of Distributed Denial of Service Attacks Carried Out by Botnets in Software-Defined Networks

Jaime Tamayo[1], Lorena Isabel Barona López and Ángel Leonardo Valdivieso Caraguay[1]

[1] Departamento de Informática y Ciencias de la Computación (DICC),
Escuela Politécnica Nacional, Ladrón de Guevara E11-253, Quito - Ecuador
`{jaime.tamayo01,lorena.barona, angel.valdivieso }@epn.edu.ec`

**Abstract.** Recent years witnessed a surge in network traffic due to the emergence of new online services, causing periodic saturation and complexity problems. Additionally, the growing number of IoT devices further compounds the problem. Software Defined Network (SDN) is a new architecture which offers innovative advantages that help to reduce saturation problems. Despite its benefits, SDNs not only can be affected by traditional attacks but also introduce new security challenges. In this context, Distributed Denial of Service (DDoS) is one of the most important attacks that can damage an SDN network's normal operation. Furthermore, if these attacks are executed using botnets, they can use thousands of compromised devices to disrupt critical online services. This paper proposes a framework for detecting DDoS attacks generated by a group of botnets in an SDN network. The framework is implemented using open-source tools such as Mininet and OpenDaylight and tested in a centralized network topology using BYOB and SNORT. The results demonstrate real-time attack identification by implementing an intrusion detection mechanism in the victim client. Our proposed solution offers quick and effective detection of DDoS attacks in SDN networks. The framework can successfully differentiate the type of attack with high accuracy in a short time.

## 1 Introduction

Software-defined networks (SDNs) offer a novel network model that differs from traditional networks and provides several advantages, such as increased flexibility, scalability, and centralized management. Communication networks shift from a proprietary infrastructure to a more flexible, open, and programmable one [1]. The SDN architecture separates de data and control planes in network devices and centralizes the control logic of the different devices (switches), which only forward the traffic. That is, the controller centrally makes the decisions on the data paths in the network [2]. The controller can be programmed according to the particular needs of each user. On its part, the switches follow the instructions provided by the controller. In addition, SDNs are more scalable than traditional networks [3].

The volume of traffic circulating on public and private networks has increased thanks to the emergence of new online services. For this reason, networks periodically suffer from significant saturation and complexity problems [4]. Traditional network environments are not self-configuring and, therefore, cannot adapt appropriately when new loads appear on the network. Additionally, the flexibility of network growth is reduced by presenting a vertical type of integration [3]. For conventional routers, it is effort and time-consuming to modify their behavior. In this case, administrators have to configure each device individually to change high-level network policies, often using vendor-specific commands. On the other hand, the configuration of an SDN network is more dynamic and efficient. The existence of a central controller in an SDN network makes the routing and controlling task much easier than in a traditional network [5].

Security is expected to be an important application area for SDN. Because of the SDN architecture is innovative, it brings new challenges to network security. Security problems on SDN networks can be divided into two types: attacks inherited from traditional networks and new attacks aimed explicitly at SDN networks. In this regard, one of the most recurrent attacks is the Denial of Service (DoS), which causes damage to both the users and the service provider. This type of attack can affect network functionality and result in financial and prestige losses.

DoS attacks are a problem for the proper functioning of a network since their main objective is disrupting services by limiting access to a machine or service [6]. Furthermore, there is a variation of DoS which can increase the damage on the network. An attack is considered a Distributed Denial of Service (DDoS) when the DoS attack is coordinated. The attacker identifies vulnerabilities and installs malware on multiple machines to control them. After, the controlled machines become a botnet and are commanded by a botmaster [7]. The botmaster can execute malicious actions on the infected devices [8]. If the number of compromised machines is enormous, it can take down the service of an entire web server in a very short time. For example, in February 2022, a distributed denial of service (DDoS) attack was launched against the websites of banks and the Ministry of Defense of Ukraine, after which many countries point to Russia as responsible showing that cyberspace is playing an important role in the conflict between Russia and Ukraine [9].

Figure 1 shows the classification of DDoS attacks, from which flood and amplification attacks will be selected for this paper. These attacks were selected as they are considered two of the most important attacks due to their effectiveness in exhausting the resources of the target server and affecting the availability of online services [10]. These attacks have been extensively documented by security organizations and have been the subject of numerous reports and analysis in the cybersecurity community [10].

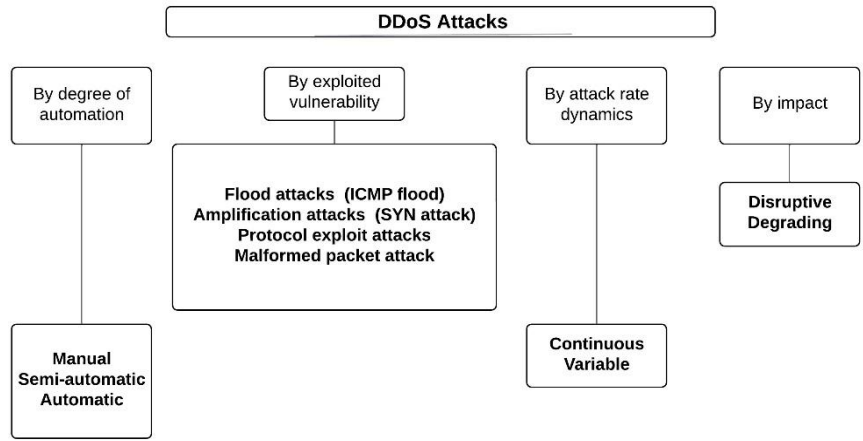

**Figure 1**. Classification of DDoS attacks

If we refer to DDoS attacks on SDN networks, they are generated by creating many flows that flood the bandwidth of the control plane, saturating the switches and the controller of an SDN network. When a controller gets compromised and experiences a DDoS attack, it can cause the entire network to fail since the controller is in charge of implementing the network logic and managing the applications and switches.

For all the above, our proposed framework aims for early detection of DDoS attacks in SDNs. The framework models the different elements who interact in the attack on SDN networks and introduce elements able to detect the attack in real time. To demonstrate the feasibility of the framework, the SDN network is created in a controlled environment using virtual machines. The botnet is created to control some users of the SDN through the malware script and turn them into zombies. The detection is performed by SNORT, which is installed on the victim's machine. Results show that the executed DDoS attacks are effectively detected shortly after starting the attack. Lastly, SDN network architecture continues to be an interesting option for ever-growing environments. Therefore, the implementation of a DDoS attack detection technique is a useful contribution to developing frameworks for the security of SDNs.

The paper is organized as follows. Section II gives an introduction and overview of SDNs and DDoS attacks carried out on this type of network, as well as the use of botnets to carry out DDoS attacks. Section III discusses the related works detecting DDoS attacks in traditional networks and SDN. Section IV discusses the implementation of the framework. The experimental results are shown in Section V; finally, Section VI shows the conclusions reached in this project and indicates studies that can be carried out in the future.

## 2 Main concepts

### 2.1 Software-Defined Networks (SDN)

SDN is a network architecture that works differently from the traditional network. This architecture is divided into three layers: application layer, control layer and infrastructure layer or data plane, as depicted in Figure *2*. The control and data planes are decoupled in SDN architecture, and the underlying network infrastructure is abstracted from upper applications [3].

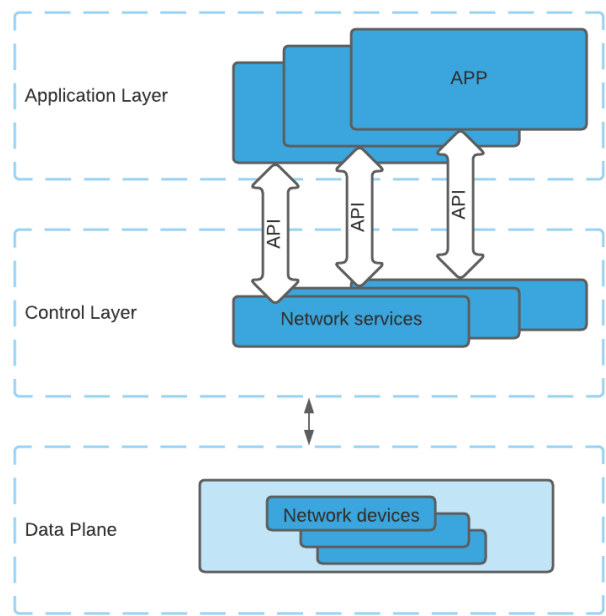

**Figure 2**. SDN architecture

The data plane contains physical or virtual switch devices, which become simple forwarding devices. The most notable example of a data plane device is the OpenFlow switch [11], [4]. An OpenFlow switch has one or more tables of packet handling rules (flow table). Depending on the rules installed by a controller application, the controller can instruct an OpenFlow switch to behave like a router, switch, firewall, or perform other roles (in general, those of a middlebox) [3].

The control layer is usually composed of logically centralized software-based controllers; they conduct the behavior of the data plane through southbound API and provide network services to upper applications through northbound API. OpenFlow protocol is SDN's first and most widely deployed protocol [1]. The OpenFlow protocol defines the communication mechanism that enables the SDN controller to interact with the data plane directly. The controller pushes packet handling rules in flow tables of OpenFlow switches. The rule matches the traffic conditions and performs specific actions, such as dropping, forwarding, and modifying traffic [1].

For its part, the application layer mainly consists of various business applications, such as virtual networks, security, and traffic engineering applications. The application plane is on the top of the SDN architecture and contains the SDN applications for various functionalities, such as policy implementation, network management, and security services. In the application plane, the SDN application can use the programmable method to submit the network behavior to the control plane [12].

### 2.2 Botnets and DDoS Attacks on SDNs

A Denial of Service (DoS) attack intends to interrupt the regular operation of the device or network and disrupt the user's activity [5]. One of the most used flood attacks is the ICMP flood [13]. The concept is to send spoofed IP packets to all hosts in the network. This amplifies the traffic and stops the processing of legitimate packets in the network. Another type of DoS attack is Distributed Denial of Service (DDoS) which occurs when multiple synchronized devices perform a DoS attack on one victim [5].

The objective of a DDoS attack is to bring down a target's services using multiple distributed sources. The idea of DDoS attacks revolves around the fact that a large number of sources distributed across various locations are used to target a victim [1]. Botnets are typically helpful for launching DDoS attacks as they are an extensive collection of compromised hosts (also called zombies). The way the bots are controlled depends on the architecture of botnet command and control mechanisms, which may be IRC, HTTP, DNS, or P2P-based. Since DDoS attacks are frequent, the focus has been to develop a proficient solution capable of effectively detecting DDoS attacks. To launch an attack, an attacker generally follows four basic steps [6]:

1) Information gathering to scan a network to find vulnerable hosts to use them later to launch an attack.
2) Compromising the hosts to install malware or malicious programs in the compromised hosts or zombies so that they can be controlled only by the attacker.
3) Launching the attack to command the zombies to send the victim several malicious flows with customizable intensities.
4) Cleaning up to remove all records or history files from memory.

Current DDoS security trends show the limitations of existing technologies, which are propagated toward SDN networks. Attackers are adopting sophisticated mechanisms to bypass traditional protection shields. More recently, software-defined networks (SDN) have emerged as a new networking paradigm with wide-scale attraction. The distinct characteristics of SDN have led to the development of many SDN-based DDoS attack detection mechanisms, as discussed in the next section.

## 3    Related Works

Some previous works have ventured into the study of detection and mitigation mechanisms of DDoS attacks on SDNs.

In [7], an intrusion detection system in SDN networks is implemented using three virtual machines: the first has the controller and the IDS, the second emulates the network domain, and the third represents an online server. The system automatically detects various DDoS attacks and notifies to the controller of the infrastructure using a framework based on RYU controller. Then, it transfers new flow rules to the network devices to restore regular network operation as fast as possible. This project implements the entire detection system in the SDN network controller, which increases the amount of work that this device performs, suggesting that the detection system should be installed as close as possible to the machine where it originates the attack.

In [11], a source-based defense mechanism against a DDoS flooding attack using botnets in SDN networks and the sample Flow (sFlow) technology is proposed. The developed detection algorithm is based on a statistical inference model. In this project, an application is developed, and its proper functioning is tested in an emulated network with real traffic. The results show that this mechanism effectively detects DDoS flooding attacks in SDN environments and identifies flows to prevent the damage of the attack from spreading beyond the target. Unlike the contribution in this paper, our work allows for the design and creation of custom botnets by using BYOB to execute DDoS attacks, whereas this study only performs a simplified attack without structuring a botnet to execute the action.

The work in [4] proposes a DDoS blocking scheme applicable to an SDN-managed network. The application that does the detection and blocking work is mounted on the POX application controller, and the communication protocol between devices is OpenFlow. The emulation shows that the POX application successfully filters the DDoS attack traffic from the legitimate traffic. The malicious traffic is blocked and safely redirects the user's traffic from the attacked server address to a new address. The project under discussion requires dedicated communication between the DDoS controller and the server or user to be protected. For its part, our proposal does not need a dedicated communication since the protection system is mounted on the user to be protected.

In [1], a framework called ProDefense is proposed. It is mainly focused on the operation of large-scale infrastructures to protect against DDoS attacks for a data center in a smart city. The framework is modular and adjusts to the requirements of several applications executed in the smart city. For instance, a Traffic Control System requires an extremely agile security solution that can trigger the mitigation system immediately and generate security alerts foreseeing malicious behavior before reaching the threshold. The security solution must also monitor network traffic trends and predict the attack [1]. While this work designs the detection of several DDoS attacks based on application requirements, our work focuses on the flooding and amplification categories, as shown in **Error! Reference source not found.**. The detection system is implemented in the victim's machine.

Other studies have been done regarding the various DDoS attack detection methods. In [14], [15], and [16], some research has been carried out using the entropy technique. The entropy technique relates to network characteristic alterations to detect anomalous network activities and identify a possible DDoS attack. In [17] and [18], authors use machine learning techniques such as Bayesian networks, SOM (Self-organizing Map), or fuzzy logic to identify the presence of anomalies. The mentioned techniques take into account various network characteristics as well as traffic analysis to achieve the detection of possible DDoS attacks.

The article presented in [19] uses the pattern analysis technique that assumes that attackers exhibit similar behavior, such as sending the same type of malicious packet or performing the same scanning action within the network. These malicious activities are detected and identified as part of a DDoS attack. This shows that the malicious behavior patterns that occur in traditional networks are helpful for the detection of future attacks aimed at SDN networks. For its part, studies shown in [20] and [21] present the connection rate technique, defined as an indicator of the number of connections made within a specific time window. If this number of connections is exceeded, it can be considered a DDoS attack. In our study, a list of rules is used instead of an indicator, where if the conditions set in the rule are met, a notification of a possible DDoS attack is launched.

In [22], a system of sensors that monitor the network and a set of correlation functions in the Open vSwitches (OVS) are proposed. When a monitor sends an alert, the correlator analyzes it to see if it matches any attack signature. If it does, the monitor, controller, and correlator take action to mitigate the impact of the attack. Our proposed project tries to detect a DDoS attack by using the SNORT IDS without the use of added monitors as part of the SDN network architecture. In [23], authors use SNORT and OpenFlow to detect and mitigate DDoS attacks in real time by modifying traditional network configurations in a cloud environment. Our project uses the SNORT IDS and the OpenFlow communications protocol to detect DDoS attacks on an SDN network architecture rather than on a traditional network.

In [24] an SDN network with an ONOS controller is proposed to detect through SNORT a TCP-SYN-based attack generated from Hping3. The main difference with this project is that the SNORT rules are implemented in the ONOS controller of the SDN network, while in our project the rules are implemented in the victim's computer. In this way, the job of detecting any possible DDoS attack is assigned to the victim machine itself and not as an additional job for the controller. Also, our project compares two known DDoS attacks for detection (SYN flood and ICMP flood) as opposed to this proposed project which only addresses detection of a single attack (TCP-SYN-based). The controller used in [24] uses the ONOS platform as opposed to ours which uses the OpenDaylight platform.

In [25], the authors discuss the limitations of traditional threshold-based methods to detect and mitigate DDoS attacks in SDN networks and explore the potential of machine learning and deep learning techniques to improve detection accuracy. The authors review various existing solutions and data sets and propose a new deep learning-based approach that uses SDN-specific features and two feature selection methods to identify relevant features for DDoS attack detection. The proposed approach is tested on three data sets and achieves high accuracy rates compared to traditional machine learning techniques. The authors highlight the need for more up-to-date and diverse data sets for future research in this area.

The main difference is that our project uses the SNORT tool for detection of DDoS attacks on a simulated SDN network with Mininet and OpenDaylight, while the project discussed in [25] proposes an approach based on deep learning to improve the detection accuracy of DDoS attacks in SDN networks. In addition, our project focuses on the detection of SYN flood and ICMP flood attacks targeting a specific user in the SDN network, while the project discussed in [25] evaluates the detection accuracy on three different data sets and proposes a new approach based on deep learning for the detection of DDoS attacks in SDN networks.

In [26] DDoS attacks against centralized controllers in SDNs are detected using the Mininet emulation tool. Experiments were performed using different penetration tools to launch DDoS attacks and SNORT was integrated for detection. The results showed that ODL and ONOS controllers are vulnerable to DDoS attacks and that detection time is directly proportional to the number of network devices and the amount of traffic bombarded. Rules were created in SNORT for the detection of DDoS attacks and the implementation of a DDoS detection system using SNORT IDS in ODL and ONOS controllers in SDN is discussed. System performance was evaluated using various scenarios with different numbers of hosts, switches, and traffic, and it was found that ODL detected DDoS attacks faster than ONOS.

In [26] the implementation of its detection system is not done in the victim client, but in the controllers, which increases the response time between the attack and the detection. All the previously analyzed projects are summarized in Table 1:

**Table 1**. Related works regarding detection of SDN attacks

| Ref. | Scope | Algorithm | Experiments | Results | Weaknesses |
|---|---|---|---|---|---|
| [1] | ProDefense, SDN, DDoS Attacks | ProDefense Framework for smart city data center. | Only theoretical analysis | None | Controllers without protection against directed attacks. |
| [4] | DDoS Attacks, POX controller, botnets | Flow counter, detects and deletes bots | SDN mounted on mininet with several botnets | Attack blocked after several seconds of attack initiation | Communication between the DDoS blocking application and the server need to be protected. |
| [7] | Flow rules modification | Change on flow rules after detection of DDoS attacks | Three scenarios with different types of DDoS attacks | Average attack reaction time of 3 seconds | Detection system implemented in the controller |
| [11] | sample Flow (sFlow) technology | Statistical inference model | Emulated network with real traffic | Detects DDoS flooding attacks. | Only performs a simplified attack / no botnets used. |
| [14], [15], [16] | Network characteristic alterations. | Entropy technique | Performance tests to validate the scalability and the effectiveness of sFlow-based approach | Improved sFlow-based mechanism. | The entire flow table does not scale for high network traffic environments. |
| [19] | Malware detection for mobile devices using SDN. | Identifying suspicious network activities through real-time traffic analysis | Local testbed and GENI infrastructure | Feasibility approach. | Not all types of malware are analyzed. |
| [20], [21] | Hybrid approach to detecting scanning worms | FRESCO / Sequential hypothesis testing and connection rate limiting | Two modules with malicious scanner / redirect all the scanner's flow into a remote honeynet | FRESCO with over 90% fewer lines of code. / Successfully restricts the number of scans / highly effective. | Ensure that worms are quickly identified with an attractively low false alarm rate |
| [22] | Monitorization of SDN and its controller | Selectively inspecting network packets on demand | Detection and mitigation of TCP SYN flood attacks on GENI. | Scalable to process high volume of traffic and large scale attacks | The project uses extra monitors to achieve a right inspection of packets on demand. |
| [23] | SNORT and OpenFlow to detect DDoS attacks | Modifying network configurations in a cloud environment | Simplified testing environment including two cloud servers with. | Variation of the performance of SnortFlow agent in different scenarios. | The whole project is mounted in a traditional network, not an SDN. |

| | | | | |
|---|---|---|---|---|
| [24] | SDN network with SNORT detection | Detection of TCP-SYN-based attacks | Creation of a network analyzed with Wireshark and implemented attack detection rules with SNORT | | Increases the load on controller processing. |
| [25] | Limitations of traditional threshold-based methods. | NFDLM, a lightweight and optimized Artificial Neural Network | Design of four different models where two are based on ANN and the other two are based on LSTM to detect the attack types of DDoS. | The detection performance achieves approximately 99% accuracy for detection. | It does not focus on specific DDoS attacks but generalizes them. |
| [26] | DDoS attacks against centralized controllers. | Integration of SNORT for detection of DDoS attacks | Five different network scenarios are considered. The number of hosts, switches and data packets vary. | ODL and ONOS controllers are vulnerable to DDoS attacks | Attacks are not generated using botnets. |

As previously discussed, this article tries to overcome some of the limitations analyzed. For this reason, this proposal intends to implement a detection method mounted on the user or server victim of the DDoS attack, avoiding overprocessing in the controller. The DDoS attacks used in our project for its detection are ICMP flood and SYN flood because they are the most common and most used attacks.

## 4    Framework for DDoS-botnet detection

Our framework proposes a DDoS attack detection system in SDN networks. The framework takes the SDN architecture as a starting point and locates the elements that cause a DDoS attack based on botnets (zombies). To detect and avoid the attack, we introduce the Detection System module which analyzes the traffic circulating in the network (see Figure 3).

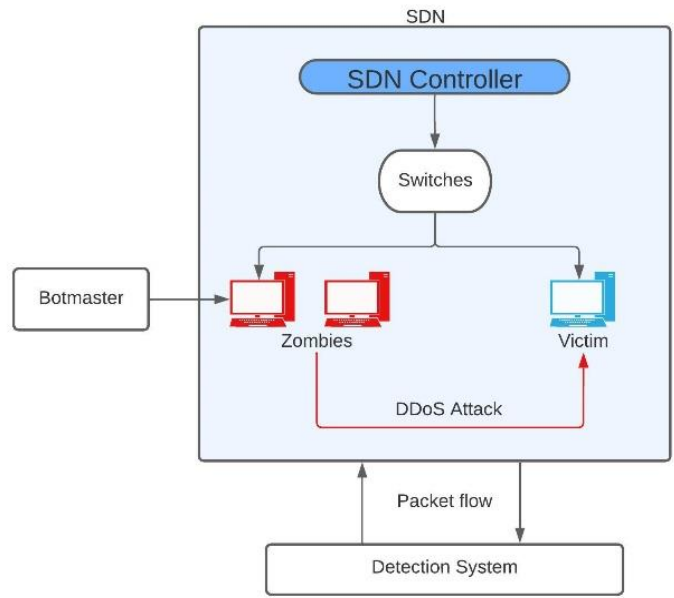

**Figure 3.** Framework for DDoS-botnet detection in SDN.

The main components of the proposed framework are:

- **SDN Infrastructure:** The SDN infrastructure contains the network hardware devices, the controller, and the users. A group of them will be transformed into zombies and will be in charge of executing the DDoS attack on a victim user.
- **SDN Controller:** It is in charge of managing the flow of data within the SDN network. In addition, the controller has the topology information and controls the configuration of each of the switches that are part of the infrastructure. The controller receives information about the source and destination of incoming packets, runs algorithms to find the optimal path, and issues commands to switches to forward the packets to their destination.
- **Botmaster:** It is in charge of managing zombie users and issuing orders to execute the DDoS attack. The botmaster can be inside or outside the SDN network and the zombies are controlled by executing malicious scripts on the computers.
- **Detection system:** It is in charge of analyzing the packets received by the user and identifying any anomaly they may have. The identification of these anomalies is done by means of previously configured rules in the detection system.
- **DDoS Attacks:** It is a group of hosts connected to the SDN network and controlled by the botmaster which carries out different attacks directed at a user within the network. It is worth mentioning that the SDN network needs the integration between the data and control planes, so the attack on any of these elements also influences the performance of the SDN controller.

The Figure 4 shows the sequence of events in which the framework works. The infrastructure and the SDN controller are the first elements that are activated for providing the communication network. At this point, users receive the network service normally. The user mounts and activates the DDoS detection module. This way, the system analyzes all user traffic and sends alert messages if a possible DDoS attack occurs.

For its part, the attacker selects the victim and starts the attack procedure. The botmaster is mounted by means of a script in order to infect devices and turn them into zombies. Once the devices are infected, the botmaster gives the order to force zombies to start to overload the victim's network resources. At this point, the detection module analyzes the traffic in real time, detects the attack and generates alert messages in the victim user.

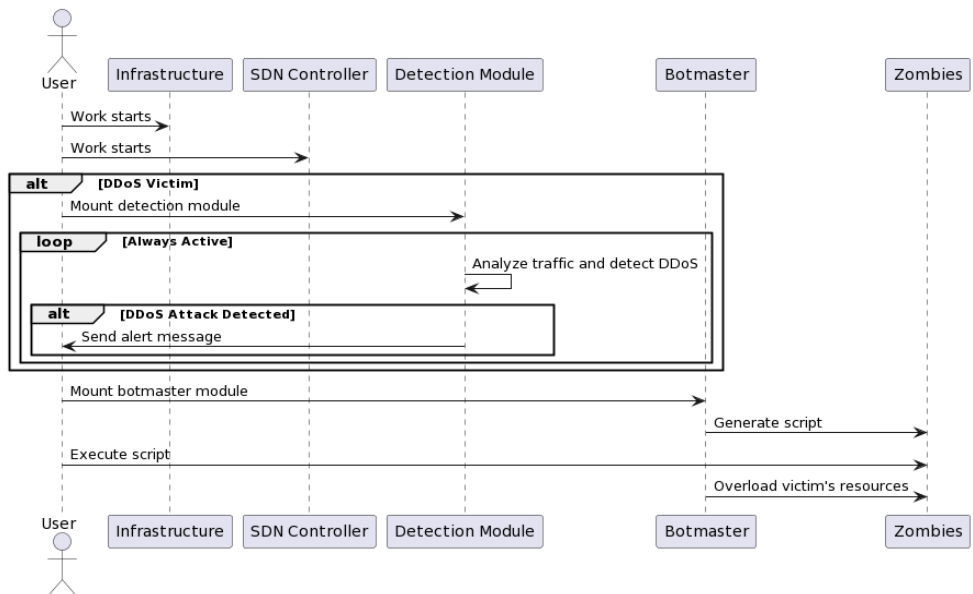

**Figure 4**. Sequence diagram of the proposed framework

## 5 Implementation

The proposed architecture is implemented used the tools described in the Figure 5. Similarly, the Table 2 describes the technical details of the used software.

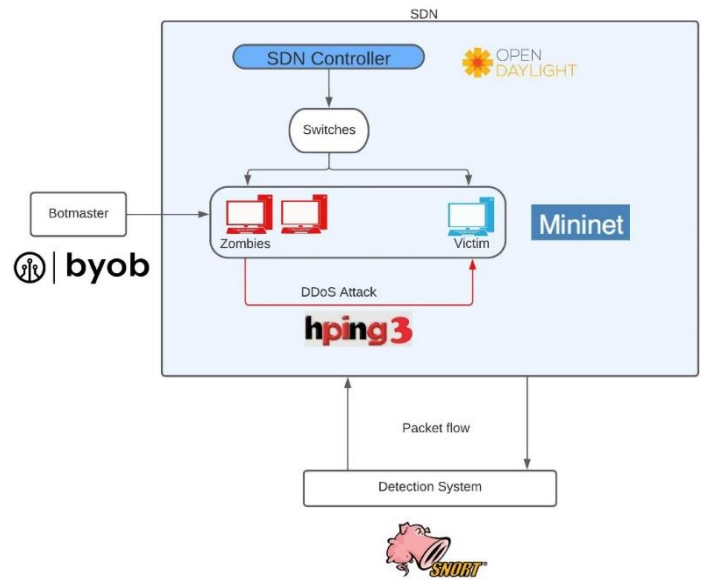

**Figure 5**. Implementation of the proposed framework

The SDN network is created from a script in Mininet; the network contains an SDN controller, a core switch, two distribution switches, and four access switches that connect twenty users. One of these users will become the victim of the DDoS attack in the two proposed versions (ICMP flood and SYN flood).

**Table 2**. Tools for detection of DDoS attacks in SDN networks

| Module | Tool | Version | Details |
|---|---|---|---|
| SDN Controller | OpenDaylight [27] | 0.13.1 | Open source<br>Handles Openflow interfaces<br>It has a GUI [28] |
| Botmaster | BYOB [29] | 2.1 | Open source.<br>Provides a command line interface (CLI).<br>Allows botnet's customization. |
| DDoS Attack | Hping3 [30] | 3 | Open source.<br>Generation of custom packets and attack flows.<br>Port and service scanning. |
| SDN infrastructure | Mininet [31] | 2.3.0d7 | Open source<br>Realistic network emulation<br>Integration with OpenFlow controllers |

| DDoS Detection | SNORT [32] | 3.1.64.0 | Open source<br>Support for multiple platforms.<br>Threat detection. |
|---|---|---|---|

The botnet attack system is created from the BYOB software that deploys a server or botmaster and is responsible for developing a Python script. When the script is executed on the hosts of the SDN network, it will become part of the botnet.

The script has the option of being created as an executable file or as a Python code file. When initialized, the botmaster takes full control of the user who is part of the SDN without the user's knowledge. The same script generated by the botmaster is used to control any number of users that are part of the network. In our project, the python script option was chosen, and it was executed on each host through its terminal.

The zombies will launch a DDoS attack on a single SDN host. The DDoS attack detection system is implemented using the SNORT tool [32]. Once installed in the virtual machine, a list of rules is created to detect ICMP flood and SYN flood attacks in a short time after the attack has been executed. Writing effective Snort rules requires a good understanding of security threats and the ability to analyze network traffic to identify potential attack patterns.

SNORT uses a simple, lightweight rules description language that is flexible and powerful. SNORT rules must completely contain a single line. The rule parser does not know how to handle rules on multiple lines. [33]

These rules are divided into two logical sections: the rule header and the rule options. The rule header contains the rule's action, protocol, source, destination IP addresses and netmasks, and the source and destination ports information. The rule option section contains alert messages and information on which parts of the packet should be inspected to determine if the rule action should be taken. [33]

### 5.1 Topology

The SDN network was created in Mininet with a script made in Python. A network containing an SDN controller, a core switch, two distribution switches, and four access switches that communicate with the OpenFlow protocol were created. The network also contains 20 users, and each access switch was connected to five of them. One of these users became the victim of the DDoS attack in the two proposed versions (ICMP flood and SYN flood), and five of these users became zombies by executing the script created by the botnet server. The SDN network topology created for the execution of experiments is shown in Figure *6* .

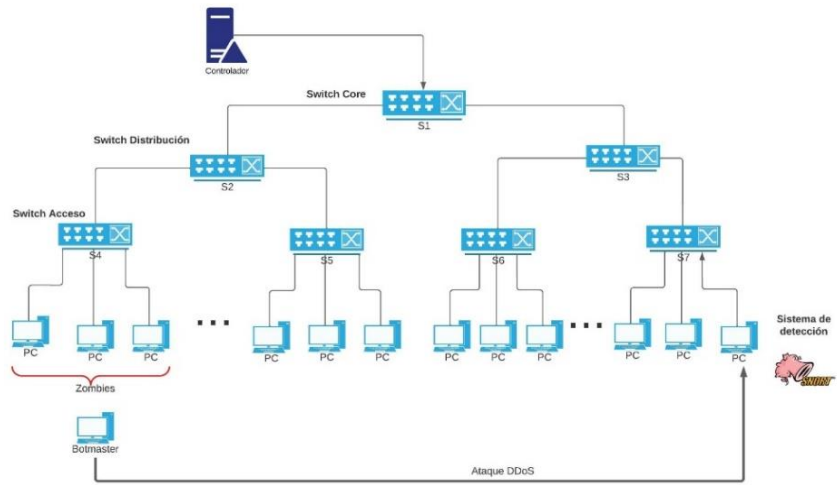

**Figure 6**. Topology of the experiments.

## 6 Experiments

The framework is evaluated using the topology mentioned in the previous section. This section presents in detail the two attack scenarios implemented and the use of the elements that generate the DDoS attack in the SDN. In the first scenario, the hosts turned into zombies are in charge of executing the ICMP Flood attack. In the second scenario, the zombies execute the SYN flood attack. Both attacks start unexpectedly targeting a network user, who has SNORT attack detection permanently active. Table 3 details the characteristics of each of the attacks. These attacks were selected as they are considered two of the most important attacks due to their effectiveness in exhausting the resources of the target server and affecting the availability of online services.

**Table 3**. Types of attacks according to the scenario

| Attack | Source | Target | Details |
| --- | --- | --- | --- |
| ICMP Flood | Zombies (Host1 through Host5). | Victim host (Host 20) | ICMP attack<br>Packet body size 1500 bytes |
| SYN Flood | Zombies (Host1 through Host5). | Victim host (Host 20) | SYN/TCP attack<br>Window size 64<br>Packet body size 100000 bytes |

# 7  Results and Discussion

## 7.1  Results

In the ICMP flood attack, the resource consumption of the victim machine is increased, both in the use of processing cores and memory consumption. Figure 7 shows the performance before and after the DDoS attack.

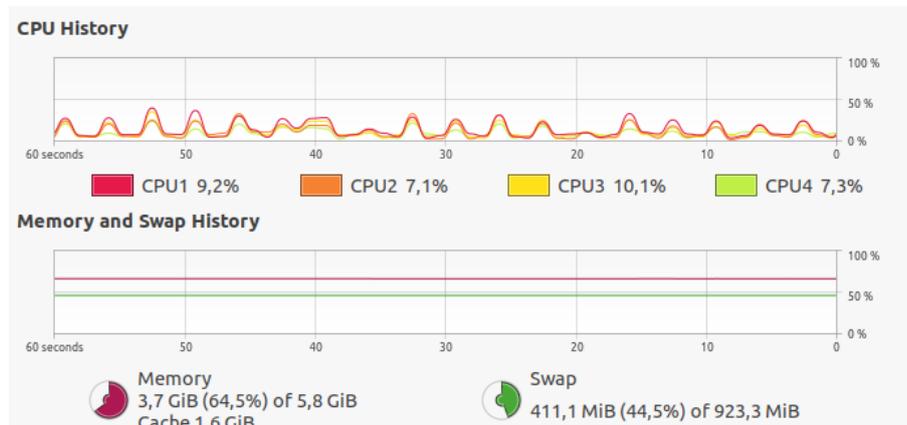

**Figure 7.** Resource consumption before and after an ICMP flood attack.

Although the resource consumption increases considerably, it does not saturate the victim's machine (a consumption of 30% on average in the 4 cores). Even so, the bombardment of ICMP packets toward the target machine is easily detected through SNORT. The time between the start of the attack and its detection is shown in Figure 8. It shows that the time range between the execution of the attack and its detection is very short. The ICMP flood attack started at 14:40:00 secs. The botmaster sent the order for the zombies to attack the victim and SNORT detected and sent alert messages of a possible DDoS attack 4 seconds after it started. Therefore, this framework is an efficient solution when detecting this type of attack.

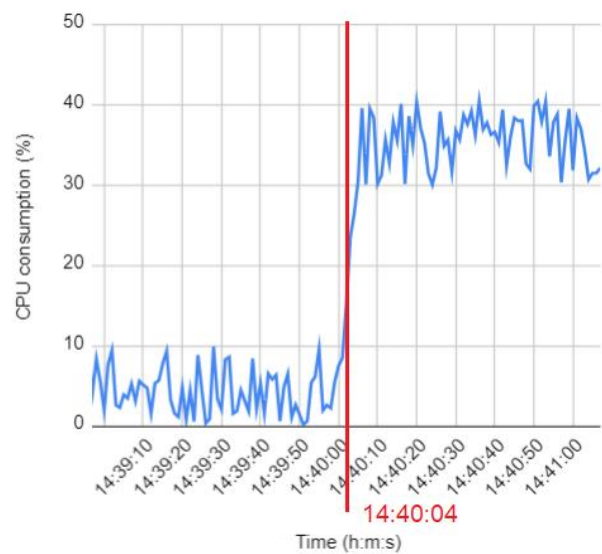

**Figure 8.** Timeline before and after the ICMP flood attack (with identification of the point at which the detection was made)

The SYN flood attack had a significant impact on the victim machine. Figure *9* shows the performance during the DDoS attack. In the SYN flood attack, unlike the ICMP flood attack, the resource consumption of the victim machine has been increased considerably. Core consumption on the victim machine was so high that most of the time, three of the four cores were above 80% consumption. Memory consumption also rose, from 15% usage to 70%.

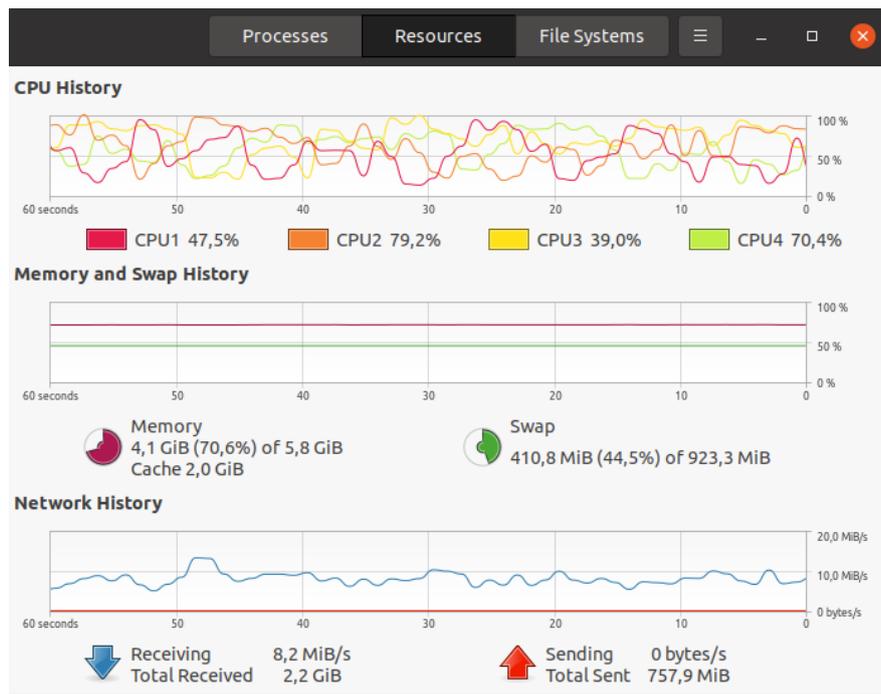

**Figure 9.** Resource consumption before and after an SYN flood attack.

Regarding detection using the SNORT tool, it showed similar results to the ICMP attack. The detection of the attack was made almost immediately. The SYN flood attack started at 14:15:05 secs. The botmaster sent the order for the zombies to attack the victim and SNORT detected and sent alert messages of a possible DDoS attack 3 seconds after it started. Figure 10 shows the time difference between the start of the DDoS attack and its detection. Even though the attack is running consuming most of the victim's resources, it can detect the attack through SNORT, and it sends a large number of alert messages right after starting the attack. Table 4 summarizes the obtained results.

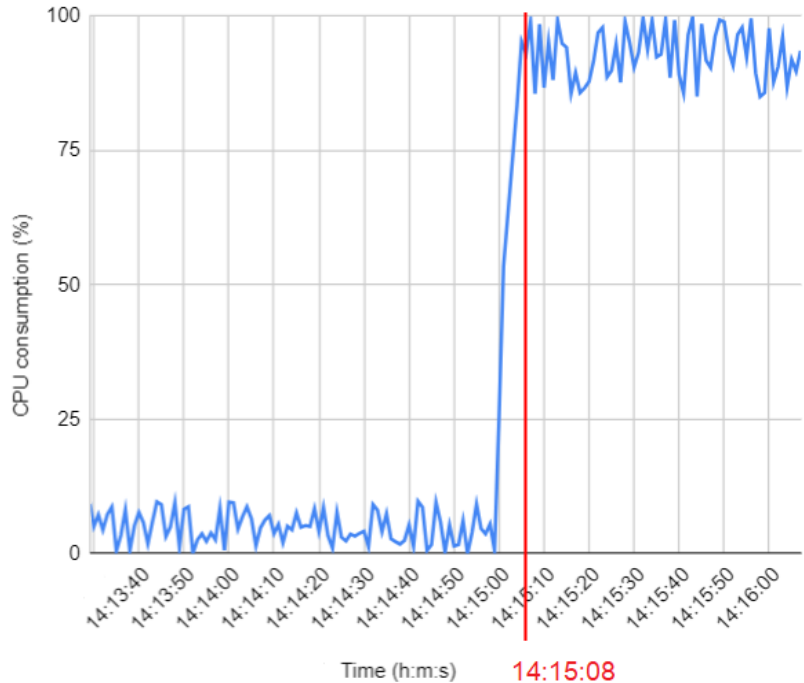

**Figure 10.** Timeline before and after the ICMP flood attack (with identification of the point at which the detection was made)

**Table 4**. DDoS attack detection results

| DDoS Attacks | Attack start time | Attack detection time | CPU consumption |
|---|---|---|---|
| ICMP Flood | 14:40:00 | 14:40:04 | 40% |
| SYN Flood | 14:15:05 | 14:15:08 | 90% |

### 7.2 Discussion

Our work focuses on the execution of DDoS attacks through two types of attacks: ICMP flood and SYN flood, since they are two of the most common attacks used in traditional networks [34]. During the execution of these two attacks, it was noticed that the SYN flood attack was detected a second faster than the ICMP flood attack. If some of the settings in the rules are changed, the difference in detection time for these same types of attacks does not change from the results shown in this section.

Resource consumption was different when running these two types of attacks. The resource impact in the ICMP flood attack was 50% lower compared to the SYN flood attack. This is because TCP packets with SYN headers bombardment were much higher than ICMP packets' bombardment. The difference in resource consumption can be seen in Figure 7 and Figure *9*.

In the SYN flood attack, it is necessary to highlight the higher consumption of resources compared to the ICMP flood attacks. These results also demonstrated the detection capacity of the SNORT tool since, seconds after the attack was carried out, the tool identified and entered the DDoS attacks in the logs.

Detection response times were slightly different. The results showed that the SYN flood attack detection time was a second faster than the ICMP flood attack detection time. However, in both cases, the detection has been successful. Even when the SYN flood attack resulted in high resource consumption for the victim, the victim managed to detect the attack almost instantly.

## 8    Conclusions

The presented architecture provides a comprehensive framework for detecting DDoS attacks generated by botnets in an SDN network. It allows for the analysis, testing, and comparison of an SDN network's behavior before and after DDoS attacks.

The implementation of the framework successfully demonstrated the detection capacity, as it quickly identified and logged DDoS attacks seconds after they were carried out. The detection response times were slightly different between the SYN flood and ICMP flood attacks, but both attacks were successfully detected.

The experiments conducted on the SDN network showed that the SYN flood attack had a more significant impact on the network's proper functioning compared to the ICMP flood attack. The resource consumption was higher in the SYN flood attack due to the bombardment of TCP packets with SYN headers. However, the victim host managed to detect the attack almost instantly, even with high resource consumption.

The framework's integration of SNORT for detecting DDoS attacks in SDN networks provides an advantage over traditional networks. Traditional networks face challenges in scaling and ensuring low false alarm rates, which SDN networks can overcome with the proposed framework.

## 9    Future work

Challenges for future work include the implementation of a DDoS attack mitigation mechanism within the framework proposed in this project. This is necessary so that actions can be taken in addition to detection, and the benefit is greater. Similarly, it is recommended the execution of experiments with a more significant number of users and, therefore, with a higher number of bots. The extension of other types of attacks additional to ICMP flood and SYN flood is also considered for future work.

## Declaration of competing interest

We declare that we have no significant competing interests including financial or non-financial, professional, or personal interests interfering with the full and objective presentation of the work described in this manuscript.

## Funding

The authors received no financial support for the research, authorship, and/or publication of this article.

## Author contributions

Jaime Orlando Tamayo Portero: Investigation, Data curation, Writing.
Lorena Isabel Barona López: Methodology.
Ángel Leonardo Valdivieso Caraguay: Conceptualization, Formal analysis.

## Data availability statement

The authors confirm that the data supporting the findings of this study are available within the article and its supplementary materials.